\documentclass[12pt,aps,prl]{revtex4}
\def\be{\begin{equation}}
\def\ee{\end{equation}}
\def\bea{\begin{eqnarray}}
\def\eea{\end{eqnarray}}

\def\fr{\frac}
\def\a{\alpha}
\def\b{\beta}
\def\d{\delta}
\def\e{\epsilon}
\def\f{\phi}
\def\g{\gamma}

\def\l{\lambda}

\def\s{\sigma}

\def\nn{\noindent}
\def\no{\nonumber}
\begin{document}
\title{On polynomial solutions of Heun equation}
\author{N. Gurappa\footnote{e-mail: nakkagurappa@yahoo.co.in} and
Prasanta K. Panigrahi\footnote{e-mail: prasanta@prl.ernet.in}}
\address{Physical Research Laboratory, Navrangpura, Ahmedabad-380
009, India}
\begin{abstract}
By making use of a recently developed method to solve linear
differential equations of arbitrary order, we find a wide class of
polynomial solutions to the Heun equation. We construct the series
solution to the Heun equation before identifying the polynomial
solutions. The Heun equation extended by the addition of a term,
$- \s/x$, is also amenable for polynomial solutions.
\end{abstract}
\maketitle

Heun equation, a second order linear differential equation having
four regular singularities, is well-known in mathematical
literature \cite{ron,maier} and has appeared in a number of
physical problems, like quasi-exactly solvable systems
\cite{khare}, higher dimensional correlated systems \cite{bhadu},
Kerr-de Sitter black holes \cite{suzuki},
Calogero-Moser-Sutherland systems \cite{take}, finite lattice
Bethe ansatz systems \cite{dor} {\it etc}. We analyze this
equation by using a newly developed method to solve linear
differential equations \cite{guru}. This approach to Heun equation
is very transparent in obtaining solutions; in particular the
polynomial solutions come out very naturally as is shown below.

 A single variable differential equation, after appropriate
 manipulation, can be cast in the form
 \be
  (F(D) + P(x, d/dx)) y(x) = 0 \quad,
  \ee
where, $D \equiv x \fr{d}{dx}$, and $F(D) = \sum_{n} a_n D^n$ is a
diagonal operator in the space of monomials, with $a_n$'s being
some parameters. $P(x,d/dx)$ can be an arbitrary polynomial
function of $x$ and $d/dx$, excluding the diagonal operators. It
can be shown that the following ansatz,

 \be
 y(x) = \sum_{m=0}^\infty (-1)^m
 \left[\fr{1}{F(D)}P(x,d/dx)\right]^m x^\l
 \ee
is a solution of the above equation, provided, $F(D) x^\l = 0$ and
the coefficient of $x^\l$ in $y(x) - x^\l$ is zero, in order to
ensure that the solution, $y(x)$, is non-singular \cite{guru}. The
fact that $D$ is diagonal in the space of monomials, $x^n$, makes
$1/F(D)$ well defined in the above expression. For explicating the
working of the present method, consider the Jacobi differential
equation \cite{gra}
 \be
 \left[(1 - x^2) \fr{d^2}{dx^2} + [ \b - \a - (\a + \b + 2) x] \fr{d}{dx}
 + n ( n + \a + \b + 1) \right] P_n^{(\a,\b)}(x) = 0 \quad,
 \ee
 where, $P_n^{(\a,\b)}(x)$'s are the Jacobi polynomials. Rewriting
 the above equation in the form of Eq. (1), one
 finds $F(D) = D^2 + (\a + \b + 1) D - n ( n + \a + \b + 1)$ and
 $P(x,d/dx) = - \fr{d^2}{dx^2} - (\b - \a) \fr{d}{dx}$. $F(D)
 x^\l = 0$ yields two solutions $\l = n$ and $\l = - ( n + \a + \b +
 1)$. From Eq. (2), it is clear that, $ \l = n$ results in a
 polynomial solution,
 \be
 P_n^{(\a,\b)}(x) = \sum_{m=0}^\infty \left[\fr{1}{(D - n) (D + \a + \b + 1)}
 \left(\fr{d^2}{dx^2} + (\b - \a) \fr{d}{dx}\right) \right]^m x^n \quad.
 \ee
 Notice that the choice of $\l = - ( n + \a + \b + 1)$ results in the
 other linearly independent solution with negative powers of $x$. We now proceed to
 study the Heun equation by using the above technique.
 The Heun equation, with four sigularities at $x = 0, 1, c$ and $\infty$,
 is given by
 \be
 \left[\fr{d^2}{dx^2} + \left(\fr{\g}{x} + \fr{\d}{x - 1} +
\fr{\e}{x - c} \right) \fr{d}{dx} + \fr{\a \b x - q}{x (x-1)(x -
c)}\right] y(x) = 0 \quad.
 \ee

Rewriting the above equation after multiplying it by $x$, one has
\bea
 \left[c x^2 \fr{d^2}{dx^2} + \g c {x} \fr{d}{dx}
 - (1+ c) x^3 \fr{d^2}{dx^2} - [(1 + c) \g + \d c + \e ]
 x^2 \fr{d}{dx} - {q}{x} \right. \no \\
 \left.  + x^4 \fr{d^2}{dx^2} + (\g + \d + \e) x^3 \fr{d}{dx} +
 \a \b x^2
 \right] y(x) = 0 \quad.
 \eea
In the above equation,
$$c x^2 \fr{d^2}{dx^2} + \g c {x} \fr{d}{dx} = F(D) = c D (D + \g -
1)$$
 and
 $$ P(x , d/dx) = A_{+1} + A_{+2}\quad, $$
 where,
 $$ A_{+1} = - (1+ c) x^3 \fr{d^2}{dx^2} - [(1 + c) \g + \d c + \e ]
 x^2 \fr{d}{dx} - {q}{x} \quad,$$
 is a degree $+1$ operator, and
 $$ A_{+2} = + x^4 \fr{d^2}{dx^2} + (\g + \d + \e) x^3 \fr{d}{dx} +
 \a \b x^2 \quad,$$ is a degree $+2$ operator, respectively.
 $d$, the degree of an operator, $O$, is defined from $[D , O] = d O$.
  Now,
 $$F(D) x^\l = 0 \quad,$$
 yields $\l = 0$ or $1 - \g$. The solutions, for the values
 of $\l = 0$ and $\l = 1 - \g$, can be written as
 \be
 y_\l(x) = \sum_{m=0}^\infty (-c)^{-m} \left[\fr{1}{D(D+\g-1)} (A_{+1} +
 A_{+2})\right]^m x^\l \quad.
 \ee
 The above is an infinite series solution to the Heun equation.
 Below, we look for the polynomial solutions.

\nn {\bf Case (i)}: Rewriting the Heun equation after dividing it
by $x$, we get,
 \bea
 \left[D^2 + (\g + \d + \e - 1) D + \a \b
 - (1+ c) x \fr{d^2}{dx^2} - [ (1 + c) \g + \d c + \e ] \fr{d}{dx}
 - \fr{q}{x} \right. \no\\
 \left. + c \fr{d^2}{dx^2} + \g c \fr{1}{x} \fr{d}{dx}\right] y(x) = 0 \quad.
 \eea
 Here,
 $$ F(D) = D^2 + (\g + \d + \e - 1) D + \a \b \quad,$$
 and
 $$\hat{P} = A_{-1} + A_{-2} \quad,$$
 where,
 $$ A_{-1} = - (1+ c) x \fr{d^2}{dx^2} - [ (1 + c) \g +
\d c + \e ] \fr{d}{dx} - \fr{q}{x} \quad,$$
 and
 $$A_{-2} = + c \fr{d^2}{dx^2} + \g c \fr{1}{x} \fr{d}{dx} \quad.$$
 The condition, $F(D) x^\l = 0$,
 after imposing the constraint $\g + \d + \e = \a + \b +
 1$, gives $\l = - \a$ or $- \b$. In order to have polynomial solutions,
 one needs to impose $q = 0$ in $A_{-1}$ and furthermore, either $- \a$ or $-
 \b$ must be a positive integer, say $n$. For the sake of illustration,
 we choose $- \a = n$. Then, the polynomial solutions to
 the Heun equation, $y_n(x)$, can be written as,
 \be
 y_n(x) = \sum_{k=0}^\infty (-1)^n \left[\fr{1}{(D - n) (D + \b)}
 (A_{-1} + A_{-2})\right]^k x^{n} \quad.
 \ee
\nn {\bf Case(ii)}: By performing a similarity transformation on
Eq.(7) by $x^{(1 - \g)}$ and writing $y(x) = x^{1 - \g} \f_n(x)$,
the choice of  $q = (\d c + \e) (\g - 1)$ yields,
 \bea
 \left[D^2 + (1 - \g + \d + \e) D + (1 - \g) (\d + \e) + \a \b
 \right.\no\\
 \left. - (1 + c) x \fr{d^2}{dx^2} - [ (1 + c) (2 - \g) + \d
 c + \e ] \fr{d}{dx} \right. \no\\
 \left. + c \fr{d^2}{dx^2} + \fr{(2 - \g) c}{x}
 \fr{d}{dx}\right] \f_n(x) = 0 \quad;
 \eea
 here,
$$ {F}(D) = D^2 + (1 - \g + \d + \e) D + (1 - \g) (\d + \e) + \a \b $$ and
$$ {P(x,d/dx)} = - (1 + c) x \fr{d^2}{dx^2} - [ (1 + c) (2 - \g) + \d
c + \e ] \fr{d}{dx} + c \fr{d^2}{dx^2} + \fr{(2 - \g) c}{x}
\fr{d}{dx} \quad. $$\\
 $F(D)x^\l = 0$ results in $\l = \g - 1 - \b $ or $ \g - 1 -
 \a$, after imposing the constraint $ \g + \d + \e = \a + \b + 1$.
 In order to have polynomial solutions, either $\g - 1 - \b$ or $
 \g - 1 - \a $ must be a positive integer. Let $\g - 1 - \b = n$,
 a positive integer, then the solution is
 \be
 y_n(x) = x^{(1 - \g)} \sum_{m=0}^\infty (-1)^m \left[\fr{1}{(D - n)(D - \g + 1 + \a)}
 P(x,d/dx)\right]^m x^n \quad.
 \ee

It is interesting to note that if one extends the Heun equation by
adding a term, $ - \s/x$, the above analysis holds and yields
polynomial solutions. Starting from
 \be \left[\fr{d^2}{dx^2} +
 \left(\fr{\g}{x} + \fr{\d}{x - 1} + \fr{\e}{x - c} \right)
 \fr{d}{dx} + \fr{\a \b x - q - \s/x}{x (x-1)(x - c)}\right] Y_n(x) = 0
 \quad,
 \ee
 and performing a similarity transformation by writing $Y_n(x) = x^{1 -
\g + \s} \chi_n(x)$, and by imposing the constraint $ q = (1 - \g
+ \s) [(1 + c) \s + \d c + \e]$ (instead of the usual one $\g + \d
+ \e = \a + \b + 1$), one finds that $\chi_n(x)$ obeys the
following differential equation,
 \bea
 \left[ D^2 + (2\s + 1 - \g + \d + \e) D + (1 - \g + \s) (\s + \d + \e) +
 \a \b \right.\no\\
 \left.- (1 + c) x \fr{d^2}{dx^2} - \{[2(1 + \s) - \g] (1 + c) +
 \d c + \e \} \fr{d}{dx} \right. \no\\
 \left. + c \fr{d^2}{dx^2} + \fr{[2(1 + \s) - \g]
 c}{x}\fr{d}{dx} \right] \chi_n(x) = 0 \quad.
 \eea
 In the above case,
 $$ F(D) = D^2 + (2\s + 1 - \g + \d + \e) D + (1 - \g + \s) (\s + \d + \e) +
 \a \b \quad,$$
 and the rest of the terms belong to $P(x,d/dx)$. Imposing $F(D) x^\l =
 0$ yields,
 $$ \l^2 + (2\s + 1 - \g + \d + \e) \l + (1 - \g + \s) (\s + \d + \e) +
 \a \b = 0 \quad,$$
 which has two roots, $\l_{\pm}$. In order to have polynomial
 solutions, either $\l_+$ or $\l_-$ or both must be a
 positive integer. The solutions can then be written as
 \be
 Y_{\l_\pm}(x) = x^{1 - \g + \s} \sum_{m=0}^\infty
 (-1)^m \left[\fr{1}{(D - \l_+)(D - \l_-)} P(x,d/dx)\right]^m
 x^{\l_\pm} \quad.
 \ee

 In conclusion, we analyzed the Heun equation by making use of a
 recently developed method to solve linear differential equations
 of arbitrary order. Apart from the series solution, we
 have obtained polynomial solutions to the Heun equation. Further,
 we extended the Heun equation by adding a term, $- \s/x$,
 and showed that polynomial solutions also exist in this case. The
 solutions found here may throw new light on physical
 problems involving Heun equation.

\end{document}